\documentclass[12pt]{article}

\usepackage{amsfonts,amssymb,graphics,pst-plot,amsthm}
\usepackage[fleqn]{amsmath}
\usepackage{graphicx}
\usepackage{authblk}

\title{Modeling and Simulation of the Effects of Cyclic Loading on Articular Cartilage Lesion Formation}
\author[1]{Xiayi Wang}
\author[1,2,3]{Bruce P. Ayati}
\author[3,4]{Marc J. Brouillete}
\author[5]{Jason M. Graham}
\author[3,4]{Prem S. Ramakrishnan}
\author[3,4]{James A. Martin}
\affil[1]{Program in Applied Mathematical \& Computational Sciences, University of Iowa}
\affil[2]{Department of Mathematics, University of Iowa}
\affil[3]{Department of Orthopaedics \& Rehabilitation, University of Iowa}
\affil[4]{Department of Biomedical Engineering, University of Iowa}
\affil[5]{Department of Mathematics, University of Scranton}

\date{}

\begin{document}
\maketitle

\begin{abstract}
We present a model of articular cartilage lesion formation to simulate the effects of cyclic loading. This model extends and modifies the reaction-diffusion-delay model by Graham et al.~\cite{Graham2012a} for the spread of a lesion formed though a single traumatic event.  Our model represents ``implicitly" the effects of loading, meaning through a cyclic sink term in the equations for live cells. 

Our model forms the basis for {\em in silico} studies of cartilage damage relevant to questions in osteoarthritis, for example, that may not be easily answered through {\em in vivo} or {\em in vitro} studies. 

Computational results are presented that indicate the impact of differing levels of EPO on articular cartilage lesion abatement.
\end{abstract}

\section{Introduction}

%\cite{Engfeldt1987,Helminen1992,Kiviranta1992,Kiviranta1988,Mak1987,Mow1989,Oettmeier1992,Palmoski1984,
%Parkkinen1992,Sah1989,Suh1995,Veldhuijzen1987}

Cartilage is a tissue that thrives in a mechanically-active environment and has been well established to be biologically responsive to physical stimuli. The effect of dynamic loads on articular cartilage is partly of interest because periodic changes in loading profiles are physiological ({\em e.g.,} walking, running, etc.) and pertinent to cartilage health and disease progression. 

Articular cartilage that lines the surfaces of lower extremity joints in humans is routinely exposed to dynamic contact stresses in the 1-5 megapascal range \cite{Buckwalter2006}.  At frequencies and rates commonly encountered in activities of daily life, stress levels in this range are not only tolerated by cartilage, but are necessary for long-term stability \cite{Arokoski2000}.  On the other hand, stresses much above 5 MPa, or stresses delivered outside physiologic norms of frequency or rate, can lead to progressive cartilage degeneration, a hallmark of osteoarthritis (OA) \cite{Natoli2009c}.  These observations have focused OA research on the cellular and molecular basis of cartilage mechano-responses.  Although a great deal of progress has been made in understanding short-term responses at the tissue level, it is still unclear how these processes unfold to cause the slowly developing, organ-wide  breakdown that occurs in osteoarthritic joints.  As a result, though mechanical stresses probably play a decisive role in many cases of OA, there are currently no therapies targeting biologic mechanoresponses.

While the chronic effects of over- or under-loading can be studied in {\em in vivo} models, the actual stresses on cartilage in most experimental animals are a matter of conjecture and difficult to precisely control. In contrast, stresses can be closely monitored and controlled in {\em in vitro} systems, but only short-term responses can be studied due to culture-related instability. These limitations leave a knowledge gap that is unlikely to be bridged by further experimental work. However, it may still be possible to extrapolate from short term {\em in vitro} data to OA-relevant time frames using {\em in silico} models \cite{Graham2012a,Wilson2005a}.  Here we describe advanced biomathematical models that draw on the wealth of knowledge of chondrocyte mechanotransduction to portray realistically cartilage stress responses.

Articular cartilage response to mechanical loading is viscoelastic, largely due to the interaction between the solid and fluid phases of its composition. To this end, cartilage is described as a biphasic material and is generally studied as a mixture of an elastic solid and interstitial fluid. The diffusive momentum exchange between the two phases regulates matrix deformation (via fluid exudation) when mechanical stimulus is imposed. 

In this investigation, we attempt to extend  and modify a reaction-diffusion-delay model of cartilage lesion formation \cite{Graham2012a} by adding features of the linear biphasic theory to simulate cyclic compressive loading. The governing equations of this model would be able to predict displacement of the solid matrix of the tissue (referred to as tissue strain) when a cyclic loading waveform is applied. As opposed to a single blunt impact injury (as was the case in \cite{Graham2012a} and explored more fully in \cite{Graham2013a}), the objective of this study is to simulate cartilage response to injurious cyclic compressive loading.  

Physiological cyclic loading generally produces tissue deformations of less than 20\%,  which are not considered to cause any meaningful destruction. The underlying criterion in this model is that chondrocytes die when consolidated tissue strains of large magnitudes (greater than 40\% of original tissue thickness) are induced. 

We make the following implicit modeling assumptions in our loading term about the material properties of articular cartilage: it is a composite structure with an intrinsically incompressible, porous and elastic solid phase (chondrocytes, collagen and proteoglycans); and the fluid phase is assumed to be intrinsically incompressible and inviscid.  Moreover, we assume cyclic loading of cartilage is imposed on a known region of cartilage surrounded by unloaded tissue. The loaded region is  simplified to be a radially symmetric one-dimensional case of cyclic compression via a porous filter in a confined configuration. 

\section{One-dimensional model with implicit mechanical loading}\label{implicitMechanics}
In articular cartilage, dynamic mechanical loads can stimulate biosynthetic activity. Studying the environment of chondrocytes under dynamic loading conditions can help explain this mechanic-biological phenomenon. In the model in this section, we modify a reaction-diffusion-delay model by Graham et al.~\cite{Graham2012a}.  In \cite{Graham2012a}, the lesion was formed by an initial, severe traumatic event with no further loading.  In the model in this section, we assume instead that there is no initial damage, but rather cyclic compressive loading on a small part of the cartilage. The loading is expressed through a deformation term in the system of partial differential equations, rather than through explicit mechanical terms.  

We assume circular symmetry so that the system components depend only on radius ($r$), time ($t$) and time delays ($\tau_1$, $\tau_2$).  We simulate an oscillating load on a small region near the origin ($0 \leq r \leq 0.5$ cm).  

\begin{figure}
\centering
\includegraphics[width=5in]{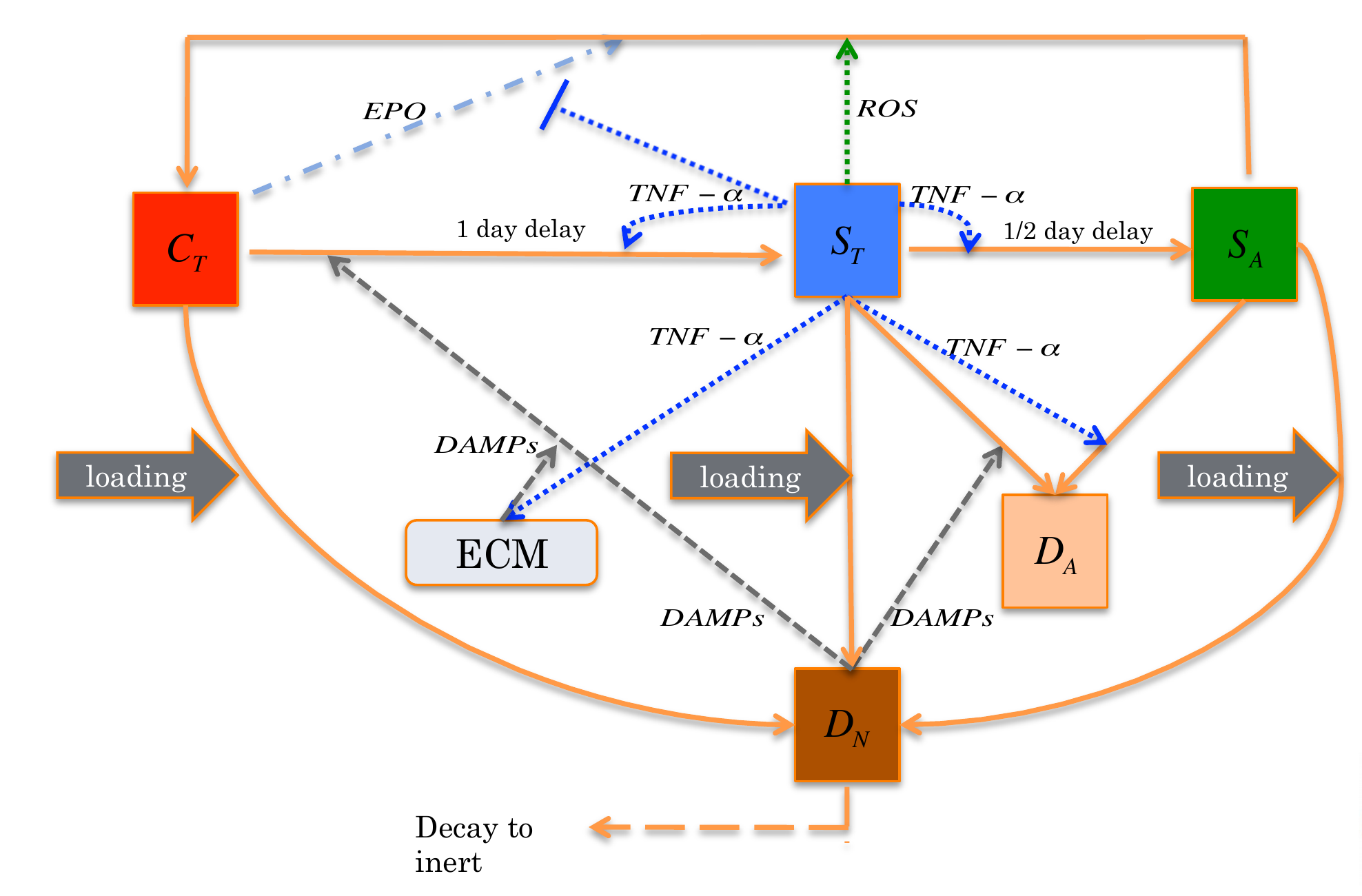}
\caption{Schmatic of the articular cartiage lesion formation process under cyclic loading.}
\label{fig:schematic}
\end{figure}

The components of our system fall into two main categories, cells and chemicals.  We also track extracellular matrix density.  A schematic of the system is presented in Figure \ref{fig:schematic}. The cellular components of our system are

\begin{itemize}
\item $C(r,t)=$ population density (cells per unit area) of healthy chondrocytes.
\item $S_T(r,t)=$ population density of ``catabolic" chondrocytes.  Catabolic chondrocytes have been
signaled by alarmins and are capable of synthesizing TNF-$\alpha$ and other cytokines associated with inflammation. Healthy cells signaled by DAMPs or TNF-$\alpha$ enter into the catabolic state and begin to synthesize TNF-$\alpha$ and produce reactive oxygen species (ROS).
\item $S_A(r,t)=$ population density of EPOR-active chondrocytes. EPOR-active chondrocytes are cells that have been signaled by TNF-$\alpha$ and express a receptor (EPOR) for EPO. It should be noted that
there is a time delay of 8-12 hours before a cell expresses the EPO receptor after being signaled to
become EPOR-active \cite{Brines2008}. 
\item $D_N(r,t)=$ population density of necrotic chondrocytes.  Necrotic (lysing) cells release DAMPs.
\item $D_A(r,t)=$ population density of apoptotic chondrocytes.  Apoptotic cells no longer play a role in the system, and are tracked explicitly to verify the conservation of cell quantities.
\end{itemize}

Since EPOR-active cells express a receptor for EPO, they may switch back to the healthy state if signaled by EPO. However, as discussed in \cite{Brines2008}, TNF-$\alpha$ limits production of EPO. Thus there is a balance between EPO and TNF-$\alpha$ that determines the spreading behavior of cartilage lesions. The catabolic and EPOR-active ``sick" classes  form the penumbra, the boundary region between the lesion and healthy tissue.  Because of the continuing role they play in the system, we explicitly track lysing necrotic cell densities ($D_N$).   Cells that have become apoptotic ($D_A$) no longer play a relevant role in the system (by definition of apoptosis). Their densities are tracked explicitly in the mathematical model for bookkeeping purposes and as a placeholder for further models where cell volume fractions may be the quantities of interest.  This differs from the model in \cite{Graham2012a} where apoptotic cells were not tracked explicitly, but were instead represented by sink terms in the equations for $S_T$ and $S_A$.

We assume chondrocytes in all states have negligible motility, although we track them explicitly in space since their densities will differ as they respond to the biochemical components of the system.

The chemical and material components of our system are

\begin{itemize}
\item $R(r,t)=$ concentration of reactive oxygen species (ROS).  ROS affects the production of EPO by healthy cells.
\item $M(r,t)=$ concentration of alarmins (DAMPs).  DAMPs signal healthy cells to enter the catabolic state, which in turn produce TNF-$\alpha$.
\item $F(r,t)=$ concentration of tumor necrosis factor alpha (TNF-$\alpha$).  TNF-$\alpha$, along with EPO, is the main driver of our system.  TNF-$\alpha$
	\begin{itemize}
	\item causes healthy cells to become catabolic,
	\item causes catabolic cells to enter the EPOR-active state \cite{Brines2008},
	\item influences apoptosis of catabolic and EPOR-active cells,
	\item causes a chain of events that leads to the degradation of extracellular matrix, which in turn increases the concentration of DAMPs (for mathematical convenience we represent these as direct effects),
	\item limits the production of EPO \cite{Brines2008}.
	\end{itemize}
\item $P(r,t)=$ concentration of erythropoietin (EPO).   EPO causes EPOR-active cells to return to the healthy state, and thus, in our model, is the check on the spread of the inflammation.
\item $U(r,t)=$ density of extracellular matrix (ECM).  ECM is degraded by TNF-$\alpha$, and in the process releases DAMPs.
\end{itemize}

The spatial dynamics of the system are governed by the diffusion of the four chemical components ($R$, $M$, $F$, $P$).  The extracellular matrix, like the chondrocytes, is assumed to have negligible motility.  

Our model equations are
\begin{subequations}\label{sys:implicit}
\begin{align}
\partial_t C(r,t)= & \alpha S_A \frac{P}{\lambda_P+P}- \beta_1 C \frac{M}{\lambda_M+M} H(P_c - P) \nonumber \\ 
& -\beta_2 C \frac{F}{\lambda_F+F}H(P_c - P)-\Gamma(\epsilon,U,r)C, \\
\partial_t S_T(r,t) = &  \beta_1 C \frac{M}{\lambda_M+M}H(P_c - P) - \beta_2 C \frac{F}{\lambda_F+F} H(P_c - P) \nonumber \\ 
&-  \gamma S_T(t-\tau_1) \frac{F(t-\tau_1)}{\lambda_F+F(t-\tau_1)} \nonumber \\
& -\nu S_T \frac{F}{\lambda_F+F} \frac{M}{\lambda_M+M} -\Gamma(\epsilon,U,r)S_T,\\
\partial_t S_A(r,t)= & \gamma S_T(t-\tau_1) \frac{F(t-\tau_1)}{\lambda_F+F(t-\tau_1)}-\alpha S_A \frac{P}{\lambda_P+P} \nonumber \\
&-\mu_{S_A} S_A \frac{F}{\lambda_F+F} -\Gamma(\epsilon,U,r)S_A,\\ 
\partial_t D_N(r,t)= & -\mu_{D_N} D_N +\mu_{S_T} S_T + \Gamma(\epsilon,U,r)(C+S_T+S_A),\\ 
\partial_t D_A(r,t)= & \mu_{S_A} S_A \frac{F}{\lambda_F+F}+\nu S_T \frac{F}{\lambda_F+F} \frac{M}{\lambda_M+M},\\
\partial_t U(r,t)= & -\delta_U U \frac{F}{\lambda_F+F}H(P_c - P) , \\
\partial_t R(r,t)= & \frac{1}{r} \partial_{r}(r D_R  R_r)-\delta_R R+\sigma_R S_T, \\ 
\partial_t M(r,t)= & \frac{1}{r} \partial_{r}(r D_M  M_r)-\delta_M M+\sigma_M D_N+\sigma_U U \frac{F}{\lambda_F+F},\\ 
\partial_t F(r,t)= & \frac{1}{r} \partial_{r}(r D_F  F_r)-\delta_F F+\sigma_F S_T,\\  
\partial_t P(r,t)= & \frac{1}{r} \partial_{r}(r D_P  P_r)-\delta_P P+\sigma_P C \frac{R(t-\tau_2)}{\lambda_R+R(t-\tau_2)} \frac{\Lambda}{\Lambda+F},  
\end{align}
\end{subequations}
for $t>0$ and $0 \leq r \leq r_m$ where $r_m = 2.5$ cm is the radius of our tissue sample.
   
The function $H(\theta)$ is the Heaviside function.  From \cite{Brines2008} we have $P_c=1$ nanomolar.

For spatial densities, we assume uniformity in the top centimeter of the cartilage so that densities per $\text{cm}^3$ are also densities per $\text{cm}^2$ on the surface.

Initial conditions are $C(r,t) = 10^5$ cells/$\text{cm}^2$, $U(r,t) = 30$ mg/$\text{cm}^2$, and $S_T(r,t)=S_A(r,t)=D_N(r,t)=D_A(r,t)=R(r,0)=M(r,0)=F(r,0)=P(r,0)=0$.  We use homogeneous Neumann boundary conditions for the chemical concentrations: $\frac{\partial R}{\partial r} \mid _{r=0}=\frac{\partial M}{\partial r} \mid _{r=0}=\frac{\partial F}{\partial r} \mid _{r=0}=\frac{\partial P}{\partial r} \mid _{r=0}=0.$

\subsection{The cyclic loading term}

Central to our cyclic loading model is the function $\Gamma(\epsilon,U,r)$ that represents the damage caused by cyclic loading. The goal of the model described in system (\ref{sys:implicit}) is to give a simple, conceptual mathematical model and simulation of the effects of cyclic loading on articular cartilage lesion formation.  The model in this section extends the model in \cite{Graham2012a} in a simple, yet still relevant way.  

To represent the effects of loading we use the function
\begin{equation}\label{eqn:GammaImplicit}
\Gamma(\epsilon,U,r)= - 24 \ln (1-0.01 p_0(e^{K_U \epsilon} - e^{20K_U}))\frac{\lambda_U}{\lambda_U+U},
\end{equation}
for $0 \leq r \leq r_l$, and $\Gamma(\epsilon,U,r) = 0$ for $r>r_l$, where $r_l$ = 0.25 cm is the radius of the region of tissue experiencing loading.  We note that $\Gamma$ is non-negative.  

The form of $\Gamma$ in (\ref{eqn:GammaImplicit}) is based on recent results on cell death as a function of equilibrium strain \cite{Brouillette2013}.  There are some limitations to using this data, even though it is the best available. The death rate is measured one hour after loading; further results are needed to build a function with respect to both strain and time.  Cells may not keep dying; death may stop at some point even if the same loading process continues.  

In section \ref{sec:simulations} we present results for different values of the strain $\epsilon$, which has unit of percent.

\subsection{Parameterization} \label{sec:parameterization}

The dependence of $\Gamma$ on ECM density $U$ means that our equation for $U$ is relevant to cell death.  We assume that ECM is only degraded by the effects of TNF-$\alpha$.  The degradation of ECM is measured by the decrease in concentration of $\text{SO}_4$.  The sulfite groups decorating the aggrecan proteins are the groups that matter -- the aggrecan protein is just an elaborate means to keep sulfates in the solid phase and in place in the matrix (so-called ``fixed charges'').  In an ``sGAG'' assay \cite{Farndale1982} there is an average of 30 g$\text{SO}_4$/L cartilage.  The molecular weight of 
$\text{SO}_4$ is 96 g/mol so that the molarity of $\text{SO}_4$ is
$$\frac{\text{30 g/L}}{\text{96 g/mol}} =  \text{0.3125 mol/L} =  \text{3125 micromolars/cm}^3.$$

To obtain the parameter $\delta_U$ for ECM decay we note that the decay rate of $\text{SO}_4$ under one nanomolar of TNF-$\alpha$ is about 16\% per week under 25 ng/ml = 1.4706 nanomolar of TNF-$\alpha$ \cite{Lu2011}.   Then the decay modulus of $\text{SO}_4$ is 
$$-\delta_U \frac{F}{\lambda_F+F} = ln(1-0.1/7)/\text{day} = -0.0144/\text{day}.$$
Using $F=1.4706$ nM and $\lambda_F = 0.5$ nM we get $\delta_U = 0.0193$/day.

For $\delta_F$ we have that the half life of TNF-$\alpha$ is around 100 hrs \cite{Wedlock1996}. So $\delta_F=-\frac{24}{100} \ln(\frac{1}{2}) =0.1664 \frac{1}{\text{day}}$. For The coefficient $\delta_P$, the half life of EPO is around 30 hrs \cite{Eckardt1989}. So $\delta_P=-\frac{24}{30} \ln(\frac{1}{2}) =0.5545\frac{1}{\text{day}}$.  For $\delta_M$, the half life of DAMPs is around 30 hrs \cite{Ito2008}. So $\delta_M=-\frac{24}{30} \ln(\frac{1}{2}) =0.5545 \frac{1}{\text{day}}$. To obtain decay rates from the experimental results in \cite{Ito2008,Wedlock1996}, we used the ``N-end rule'' \cite{Varshavsky1997}.

For the coefficient $\delta_R$, the natural half life of ROS is around 14 hrs at 0.1 nanomolar concentration. So $\delta_R=-\frac{24}{14} \ln(\frac{1}{2}) =1.1883 \frac{1}{\text{day}}$. However, under the superoxide dismute SOD, the decay rate of ROS is almost immediately. We don't know when this reaction will happen. It is hard to measure here. So we assume the coefficient $\delta_R=60$ in our model, which means the half life of ROS is smaller than 20 mins. 

To obtain the parameter $\sigma_U$ for the release of DAMPs from ECM, we assume 30 mg/$\text{cm}^3$ of ECM ({\em i.e.}, $\text{SO}_4$)  might release 10 ng/ml of DAMPs when exposed to 25 ng/ml of TNF-$
\alpha$.   To estimate the molecular weight of DAMPs we consider the weight of one species, HMGB1, which has a molecular weight of 29 kDA.  Recall the source term for DAMPs is $\sigma_U U \frac{F}{\lambda_F+F}$.  Then
\begin{align*}
\sigma_U U \frac{F}{\lambda_F+F} &=\text{change rate of DAMPs}=10 \ \text{ng}/(\text{cm}^3 \cdot \text{day}) \\ 
&=\frac{10 \times 10^{-9} \text{g /cm}^3 \cdot \text{day}}{29000 \text{g/mol}}=0.3448 \times 10^{-9} \text{mol}/(\text{L} \cdot \text{day}).
\end{align*}
Using $U = 30$ mg/$\text{cm}^3$, $F=1.4706$ nM and $\lambda_F = 0.5$ nM, we get
\begin{align*}
\sigma_U &= 0.3448 \cdot 0.7463 /30 \times 10^{-9}  \frac{\text{mol/L } \cdot \text{day}}{\text{mg/cm}^3}\\
&=0.0154 \ \frac{\text{nanomolar} \cdot \text{cm}^3}{\text{mg}\cdot \text{day}}.
\end{align*}
We need to reconcile the dimension to our model.  The behavior of the cartilage is sufficiently homogeneous at the top 1 cm layer that we can use in the model the parameter $\sigma_U = 0.0157  \ \frac{\text{nanomolar}\cdot \text{cm}^2}{\text{mg}\cdot \text{day}}$.

To obtain the parameter $\sigma_R$ for the release of ROS from catabolic cells,  we assume 1-2\% of oxygen consumed is converted to superoxides. Zhou et al.~\cite{Zhou2004} estimate the maximum oxygen consumption rate to be 10 nMoles per million cells per hour in normal conditions (5\%-21\% oxygen).  Then
$$\sigma_R = \frac{0.01 \cdot 10 \text{ nMoles}}{10^6 \text{ cells} \cdot \text{h}} = \frac{0.1 \cdot 24 \cdot 10^3}{10^6}\frac{\text{nMoles}}{\text{L}} \frac{\text{cm}^3}{\text{day} \cdot \text {cells}}.$$
Simplifying and assuming heterogeneity of the top layer, we get 
$$\sigma_R = 0.0024  \ \frac{\text{nanomolar}\cdot \text{cm}^2}{\text{cells}\cdot \text{day}}.$$

To obtain the parameter $\sigma_F$, we use that the release rate of TNF-$\alpha$ by catabolic cells is 100 pg/ml $\cdot$ per 12 hr by $5\times 10^4$ cells/ml \cite{Terada2011}.  Then
$$\sigma_F =\frac{100 \times 10^{-12} \times 10^{3} \text{g} \cdot (\text{cm}^3 /\text{L})}{5\times 10^4 \times 0.5 \times 17 \times 10^3 \text{(g/mol)} \cdot \text{cells} \cdot \text{day}}.$$
Simplifying and assuming heterogeneity of the top layer, we get
$$\sigma_F= 2.35 \cdot 10^{-7} \ \frac{\text{nanomolar}\cdot\text{cm}^2}{\text{cells}\cdot \text{day}}.$$

To obtain the parameter $\sigma_M$, we use that the release rate of HMGB1 is 3 ng/ml $\cdot$ per day by $2\times 10^5$ cells/ml \cite{Terada2011}.  Then  
$$\sigma_M=\frac{3 \times 10^{-9} \times 10^{3} \text{g} \cdot (\text{cm}^3 /\text{L})}{2\times 10^5 \times  29 \times 10^3 \text{(g/mol)} \cdot \text{cells} \cdot \text{day}}.$$
Simplifying and assuming heterogeneity of the top layer, we get
$$\sigma_M=5.17 \cdot 10^{-7} \ \frac{\text{nanomolar}\cdot\text{cm}^2}{\text{cells}\cdot \text{day}}.$$

To obtain the parameter $\sigma_P$, we use that the release rate of EPO by healthy cells is 18 ng/ml $\cdot$ per 4 days by $10^5$ cells/$\text{cm}^2$ \cite{Brines2008}.  Then
$$\sigma_P=\frac{18 \times 10^{-9} \times 10^{3} \text{g} \cdot (\text{cm}^3 /\text{L})}{ (10^5)^{\frac{3}{2}} \times 4 \times  34 \times 10^3 \text{(g/mol)} \cdot \text{cells} \cdot \text{day}}.$$
Simplifying and assuming heterogeneity of the top layer, we get
$$\sigma_P =4.2 \cdot 10^{-5} \ \frac{\text{nanomolar}\cdot\text{cm}^2}{\text{cells}\cdot \text{day}}.$$
For $\sigma_P$ we also conduct runs with a higher value corresponding to treatment that increases EPO; this is not a ``natural'' production rate.  We choose $\sigma_P  = 0.0033$, which is high enough to trigger the Heaviside functions in the model so that $P$ becomes larger than $P_c$ and shuts off the inflammation response. 

The parameters $K_U$ and $p_0$ are experimental settings \cite{Brouillette2013}.  Diffusion coefficients were obtained from measurements presented in \cite{Leddy2004}, delays from \cite{Brines2008}, and the remaining parameters were approximated.  The parameters for system (\ref{sys:implicit}) are summarized in Table 1.

\subsection{Simulation results}\label{sec:simulations}

We computed results of our system for four values of the strain, $\epsilon=0.3$, 0.4, 0.6, and 0.8, crossed with two values of the EPO production parameter: a ``low'' value of $\sigma_P = 4.2 \cdot 10^{-5}$ and a ``high'' value of $\sigma_P = 3.3 \cdot 10^{-3}$.  The low value corresponds to the parameter obtained from \cite{Brines2008} in Section \ref{sec:parameterization}, whereas the high value is set to trigger the Heaviside functions in our system and can correspond, for example, to the effects of treatment.  Extremal values at $t=10$ days of the system variables are presented in Tables 2 and 3 and provide an overview of the behavior of entire system.  We focus on simulations results most relevant to understanding the inflammation process, namely the spatial and temporal dynamics of the healthy, catabolic and EPOR-active cell populations.  

\begin{figure}
\centering
\includegraphics[width=5in]{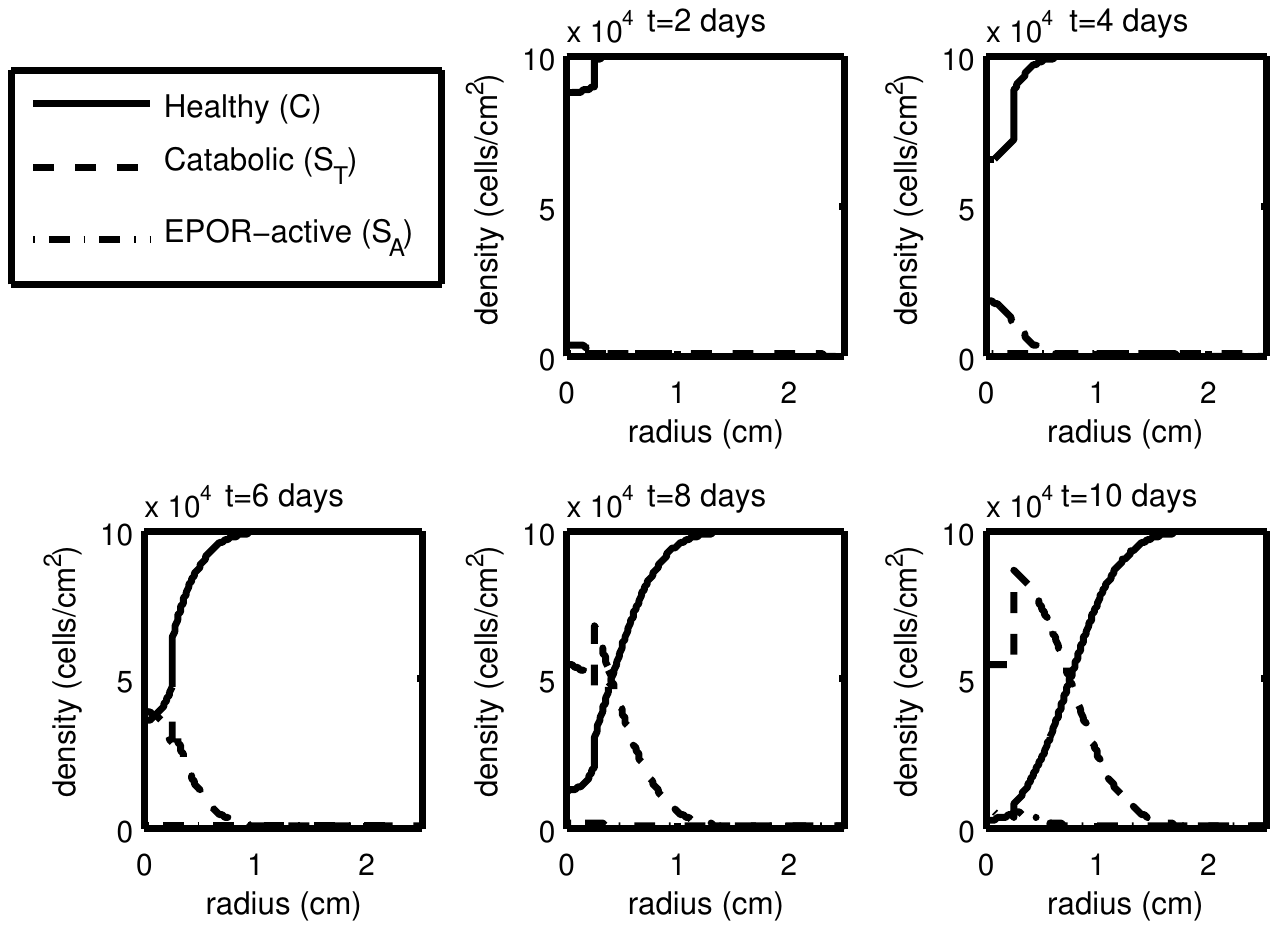}
\caption{The density of the healthy, catabolic and EPOR-active chondrocytes ($C(r,t)$, $S_T(r,t)$, $S_A(r,t)$) at $t=2,4,6,8,10$ days with $\epsilon=0.4$ and $\sigma_P = 4.2 \cdot 10^{-5}$ (light strain with low EPO production).}
\label{fig:strain40EPOlow}
\end{figure}

\begin{figure}
\centering
\includegraphics[width=5in]{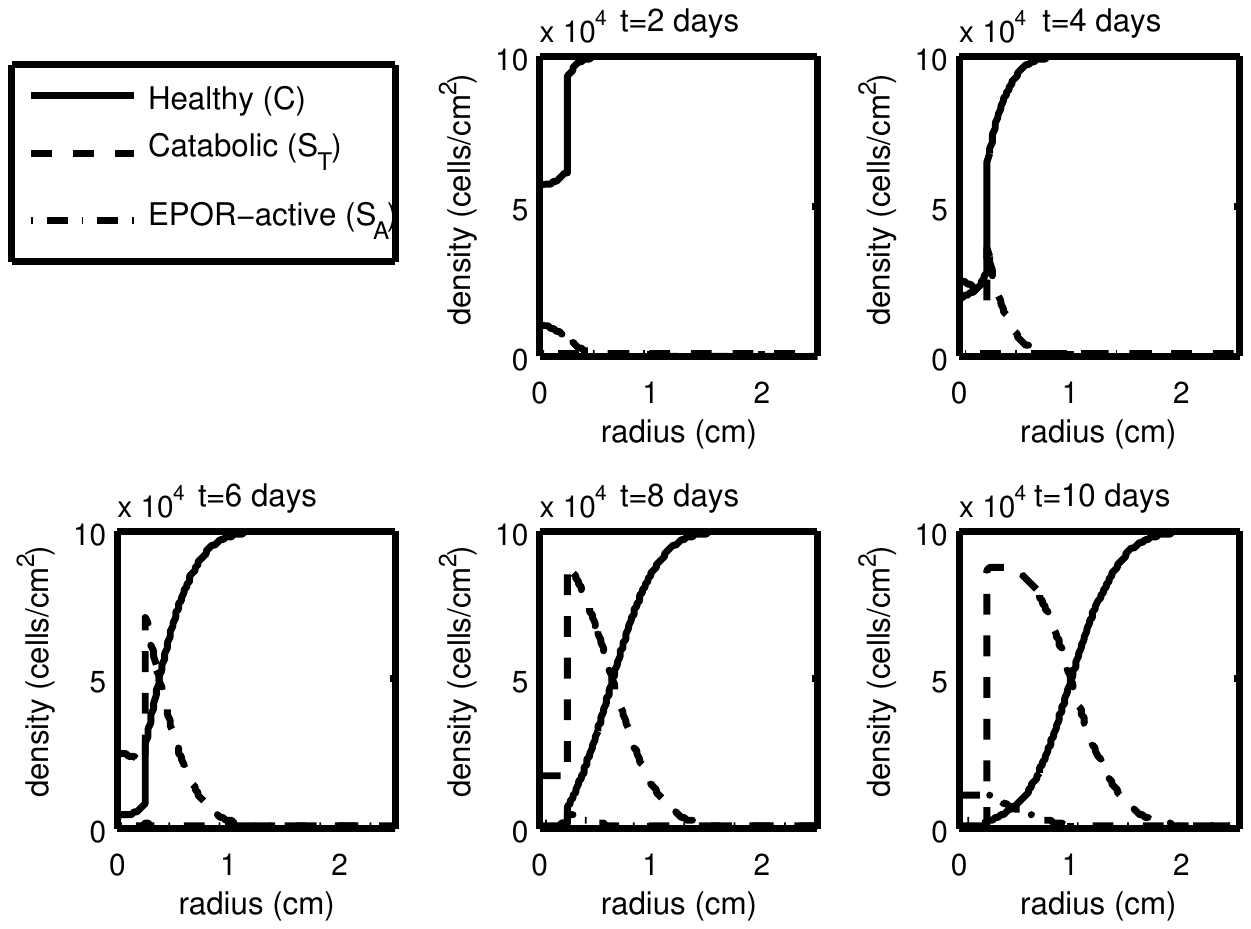}
\caption{The density of the healthy, catabolic and EPOR-active chondrocytes ($C(r,t)$, $S_T(r,t)$, $S_A(r,t)$) at $t=2,4,6,8,10$ days with $\epsilon=0.6$ and $\sigma_P = 4.2 \cdot 10^{-5}$  (medium strain with low EPO production).}
\label{fig:strain60EPOlow}
\end{figure}

\begin{figure}
\centering
\includegraphics[width=5in]{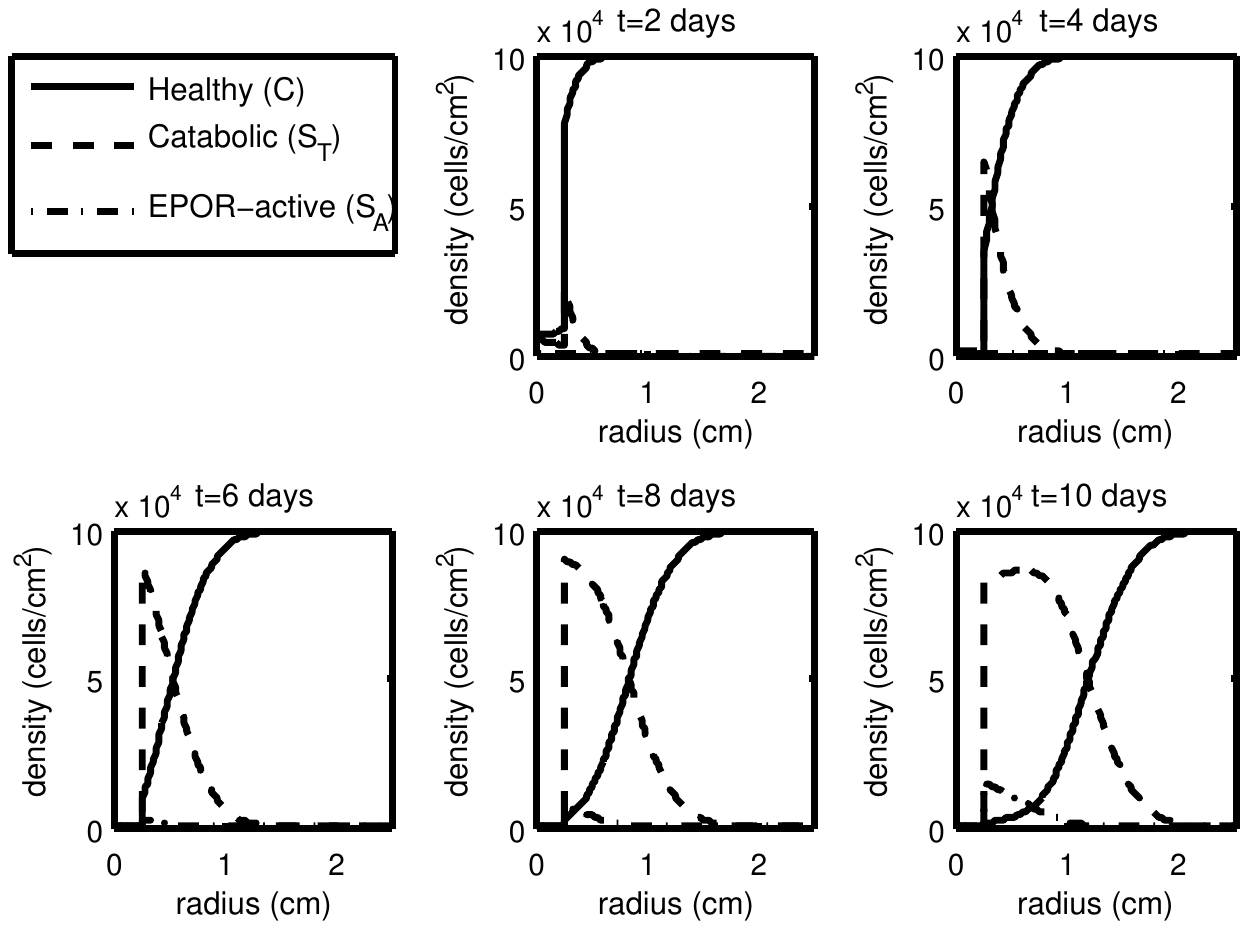}
\caption{The density of the healthy, catabolic and EPOR-active chondrocytes ($C(r,t)$, $S_T(r,t)$, $S_A(r,t)$) at $t=2,4,6,8,10$ days with $\epsilon=0.8$ and $\sigma_P = 4.2 \cdot 10^{-5}$  (heavy strain with low EPO production).}
\label{fig:strain80EPOlow}
\end{figure}

\begin{figure}
\centering
\includegraphics[width=5in]{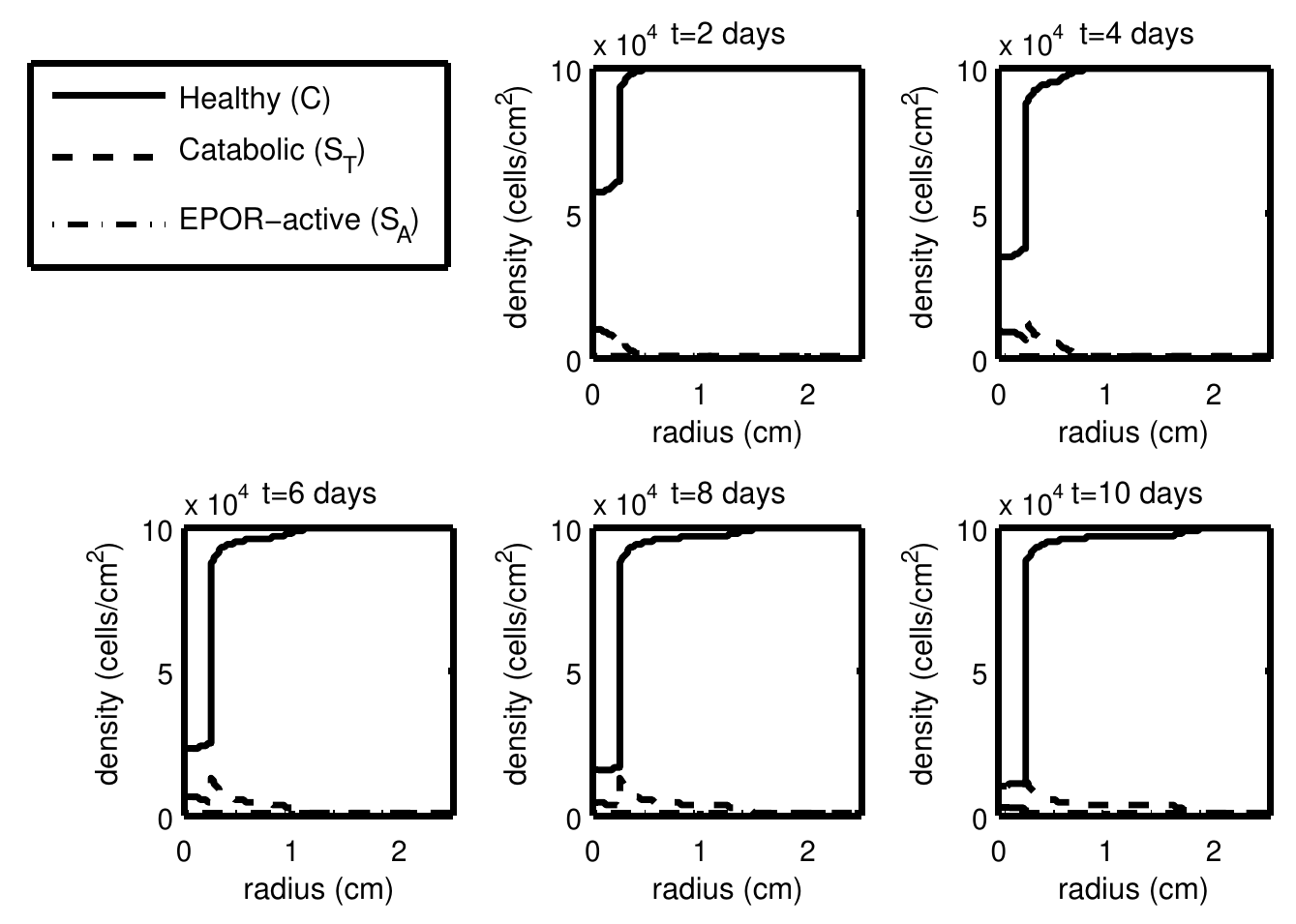}
\caption{The density of the healthy, catabolic and EPOR-active chondrocytes ($C(r,t)$, $S_T(r,t)$, $S_A(r,t)$) at $t=2,4,6,8,10$ days with $\epsilon=0.6$ and $\sigma_P = 3.3 \cdot 10^{-3}$(medium loading with high EPO production).}
\label{fig:strain60EPOhigh}
\end{figure}

To more fully understand the changes in the system due to changes in the strain, we present computational results for $\epsilon = 0.4$, 0.6, and 0.8 at low EPO production in Figures \ref{fig:strain40EPOlow}-\ref{fig:strain80EPOlow}.  To understand the effects of increasing EPO production, we present computational results for $\epsilon = 0.6$ at high EPO production, noting that the responses at other strains are similar.  Computational results for $\epsilon = 0.6$ at high EPO production are shown in Figure \ref{fig:strain60EPOhigh}.

As expected, higher levels of strain result in lower levels of healthy cells and a larger area of inflammation.  Also as expected, elevated EPO levels result in a check on the inflammation process.   The dynamics in the ''penumbra'', the region of catabolic and EPOR-active cells, is perhaps the most insightful result, showing a preponderance of catabolic over EPOR-active cells.  

The success of our model in incorporating a relatively large number of experimentally measured parameters and obtaining results matching observed inflammation response under our ``low'' EPO case validates the mechanistic assumptions that went into our model.  As such, we find from our simulation results greater confidence that the model itself is a firm foundation for a truly predictive model.  More importantly at this stage of theoretical development, the simulation results indicate that we have brought together disparate experiments and piecemeal understandings of system components and formed a more holistic understanding of articular cartilage lesion under cyclic loading.

For the ''high'' EPO case, much more work remains to be done, both in model refinement and validation.  How one increases EPO or other chemicals that act like EPO in the cartilage environment affects both model refinement and model validation.

Although perhaps of less interest to a clinician, knowing the dynamics of live cell subtypes within the penubra is an important means of validating these results. 

\subsection{Numerical methodology}

We first did a semi-discretization in space, using the radially symmetric finite difference scheme presented in Appendix C of \cite{Ayati1998}. The semi-discrete system of delay-differential equations was then solved in MATLAB by using {\tt dde23} \cite{Shampine2001}. This approach is suitable for the models in this paper and was also used to solve the model equations in \cite{Graham2012a}.  Computational convergence studies for each dependent variable in system (\ref{sys:implicit}) are shown in Tables 4-7.   The relative errors found are well within experimental or real world measurement error, are primarily due to tracking a sharp front, and there is no indication that the computed solutions differ qualitatively from the true solutions.

\section{Conclusions}

In this paper we presented biomathematical models based on observations of chondrocyte mechanotransduction.  When run in concert with a finite element program to incorporate the physical effects of mechanical compression, our new model may prove to be useful for predicting the consequences of long-term exposure to the broad range stresses experienced by cartilage in human joints.  As such the model may be useful for identifying critical stress thresholds that, over time, increase the risk for osteoarthritis.  

We have demonstrated that using delay differential equations to model the delays in the cellular responses to cytokines in articular cartilage lesion formation is a reasonable approach for models with the complexity of those in this paper.  However, as model complexity increases, the use of delay differential equations will lead to computational challenges down the line due to the need to keep in memory vast past history information.   Adding somewhat to these challenges is that the commercial off-the-shelf (COTS) solution, {\tt dde23} in Matlab, uses an explicit time integration method for the differential equations.  The stability constraints introduced by the semi-discretization in space force {\tt dde23} to take time steps much smaller than what truncation error alone would dictate.    This, in turn, exasperates the memory issues caused by the need to store the past history of the system.  Once solution may be to use a different time discretization.  However, even a ``stiff'' DDE solver for the programming environment R was found not to be appreciably faster \cite{Soetaert2012}.

The limitations of using delay differential equations and the computational methods for them may be ameliorated by converting the delay into a physiological property of individual cells, namely time since exposure to the relevant cytokine.   The partial differential equations for the cell populations that result from this approach include an additional independent variable and a derivative term for ``age'' structure.  Essentially, one would be taking the equivalent of a non-Markov process and converting it into the equivalent of a Markov process.  Although the addition of an extra dimension to the problem may seem like an unwanted complication, the reality is that the delays in a delay-differential equation require the retention of time histories of the system that can quickly become much larger to store in memory than the extra age dimension.  This difference in cost is even more pronounced if we use highly efficient methods for age- and space-structured problems \cite{Ayati2000,Ayati2007,Ayati2002,Ayati2009}.   These methods have a history of effective use in the modeling and simulation of biofilms \cite{Ayati2011,Ayati2007a,Ayati2012,Klapper2007}, avascular tumor invasion \cite{Ayati2005}, and {\em Proteus mirabilis} swarm colony development \cite{Ayati2006,Ayati2007b,Ayati2009a}.    Using age structure to represent the delays in cellular responses to cytokines is an approach suggested by our experiences with the simulations in this paper.

Despite the challenges the use of delay terms presents for computational studies with greater complexity, the use of delays in the first efforts in \cite{Graham2012a} and in this paper is warranted; we have the tautology that delay terms are more easily understood to represent delays.  Moreover, some modelers may find the models with delays are better suited for inclusion in their own efforts. Other modelers may find that the greater flexibility inherent in using ``age", or some other physiologically structured variable, makes that approach preferable.

\section*{Acknowledgments}
BPA was partially supported by the NSF under award DMS-0914514. XW, BPA, MJB, PSR, and JAM were partially supported by NIAMS grant \#1 P50 AR055533. MJB was supported by a Merit Review Award from the Department of Veterans Affairs. 

\bibliographystyle{plain}	
\bibliography{library}

\begin{thebibliography}{10}

\bibitem{Arokoski2000}
J~P~A Arokoski, J~S Jurvelin, U~V\"{a}\"{a}t\"{a}inen, and H~J Helminen.
\newblock {Normal and pathological adaptations of articular cartilage}.
\newblock {\em Scandinavian Journal of Medicine and Science in Sports},
  10(4):186--198, 2000.

\bibitem{Ayati1998}
Bruce~P Ayati.
\newblock {Methods for Computational Population Dynamics}.
\newblock {\em University of Chicago Department of Mathematics Doctoral
  Dissertation}, 1998.

\bibitem{Ayati2000}
Bruce~P. Ayati.
\newblock {A variable time step method for an age-dependent population model
  with nonlinear diffusion}.
\newblock {\em SIAM Journal on Numerical Analysis}, 37(5):1571--1589, 2000.

\bibitem{Ayati2006}
Bruce~P. Ayati.
\newblock {A structured-population model of Proteus mirabilis swarm-colony
  development}.
\newblock {\em J. Math. Biol.}, 52(1):93--114, 2006.

\bibitem{Ayati2007}
Bruce~P. Ayati.
\newblock {Modeling and Simulation of Age- and Space-Structured Biological
  Systems}.
\newblock In Kazem Mahdavi, Rebecca Culshaw, and John Boucher, editors, {\em
  Current Developments in Mathematical Biology}, pages 107--130. World
  Scientific Publishing, June 2007.

\bibitem{Ayati2007b}
Bruce~P. Ayati.
\newblock {Modeling the role of the cell cycle in regulating Proteus mirabilis
  swarm-colony development}.
\newblock {\em Applied Mathematics Letters}, 20(8):913--918, 2007.

\bibitem{Ayati2009a}
Bruce~P. Ayati.
\newblock {A comparison of the dynamics of the structured cell population in
  virtual and experimental Proteus mirabilis swarm colonies}.
\newblock {\em Applied Numerical Mathematics}, 59(3-4):487--494, March 2009.

\bibitem{Ayati2011}
Bruce~P Ayati.
\newblock {Microbial dormancy in batch cultures as a function of
  substrate-dependent mortality.}
\newblock {\em Journal of theoretical biology}, 293:34--40, October 2011.

\bibitem{Ayati2002}
Bruce~P. Ayati and Todd~F. Dupont.
\newblock {Galerkin methods in age and space for a population model with
  nonlinear diffusion}.
\newblock {\em SIAM Journal on Numerical Analysis}, 40(3):1064--1076, 2002.

\bibitem{Ayati2009}
Bruce~P. Ayati and Todd~F. Dupont.
\newblock {Mollified birth in natural-age-grid Galerkin methods for
  age-structured biological systems}.
\newblock {\em Nonlinearity}, 22(8):1983--1995, July 2009.

\bibitem{Ayati2007a}
Bruce~P. Ayati and Isaac Klapper.
\newblock {A Multiscale Model of Biofilm as a Senescence-structured Fluid}.
\newblock {\em Multiscale Model. Simul.}, 6(2):347--365, May 2007.

\bibitem{Ayati2012}
Bruce~P Ayati and Isaac Klapper.
\newblock {Models of microbial dormancy in biofilms and planktonic cultures}.
\newblock {\em Commun. Math. Sci.}, 10(2):493--511, 2012.

\bibitem{Ayati2005}
Bruce~P. Ayati, Glenn~F. Webb, and Alexander R.~A. Anderson.
\newblock {Computational Methods and Results for Structured Multiscale Models
  of Tumor Invasion}.
\newblock {\em Multiscale Model. Simul.}, 5(1):1--20, March 2006.

\bibitem{Brines2008}
M~Brines and A~Cerami.
\newblock {Erythropoietin-mediated tissue protection: reducing collateral
  damage from the primary injury response}.
\newblock {\em Journal of Internal Medicine}, 264(5):405--432, November 2008.

\bibitem{Brouillette2013}
M~J Brouillette, P~S Ramakrishnan, V~M Wagner, E~E Sauter, B~J Journot, T~O
  McKinley, and J~A Martin.
\newblock {Strain-dependent oxidant release in articular cartilage originates
  from mitochondria.}
\newblock {\em Biomechanics and modeling in mechanobiology}, July 2013.

\bibitem{Buckwalter2006}
J.~A. Buckwalter, J.~A. Martin, and T.~D. Brown.
\newblock {Perspectives on chondrocyte mechanobiology and osteoarthritis}.
\newblock {\em Biorheology}, 43(3):603--609, 2006.

\bibitem{Eckardt1989}
Kai-Uwe Eckardt, Urs Boutellier, Armin Kurtz, Michael Schopen, Erwin~A. Koller,
  and Christian Bauer.
\newblock {Rate of erythropoietin formation in humans in response to acute
  hypobaric hypoxia}.
\newblock {\em Journal of applied physiology}, 66(4):1785--8, April 1989.

\bibitem{Farndale1982}
R~W Farndale and {C A}.
\newblock {A direct spectrophotometric microassay for sulfated
  glycosaminoglycans in cartilage cultures}.
\newblock {\em Connective tissue research}, 9(4):247--248, 1982.

\bibitem{Graham2013a}
Jason~M Graham.
\newblock {A Measure of Control for Secondary Cytokine-Induced Injury of
  Articular Cartilage : A Computational Study}.
\newblock {\em Applied Mathematics and Computation}, to appear:1--31, 2013.

\bibitem{Graham2012a}
Jason~M. Graham, Bruce~P. Ayati, Lei Ding, Prem~S. Ramakrishnan, and James~A.
  Martin.
\newblock {Reaction-Diffusion-Delay Model for EPO/TNF-$\alpha$ Interaction in
  Articular Cartilage Lesion Abatement}.
\newblock {\em Biology Direct}, 7(1):9, January 2012.

\bibitem{Ito2008}
Takashi Ito, Ko-ichi Kawahara, Kohji Okamoto, Shingo Yamada, Minetsugu Yasuda,
  Hitoshi Imaizumi, Yuko Nawa, Xiaojie Meng, Binita Shrestha, Teruto
  Hashiguchi, and Ikuro Maruyama.
\newblock {Proteolytic cleavage of high mobility group box 1 protein by
  thrombin-thrombomodulin complexes}.
\newblock {\em Arteriosclerosis, thrombosis, and vascular biology},
  28(10):1825--30, October 2008.

\bibitem{Klapper2007}
Isaac Klapper, Peter Gilbert, Bruce~P. Ayati, Jack Dockery, and Philip~S.
  Stewart.
\newblock {Senescence can explain microbial persistence.}
\newblock {\em Microbiology}, 153(11):3623--3630, November 2007.

\bibitem{Leddy2004}
Holly~A Leddy, Hani~A Awad, and Farshid Guilak.
\newblock {Molecular diffusion in tissue-engineered cartilage constructs:
  effects of scaffold material, time, and culture conditions}.
\newblock {\em Journal of Biomedical Materials Research Part B: Applied
  Biomaterials}, 70B(2):397--406, August 2004.

\bibitem{Lu2011}
Yihong C~S Lu, Christopher~H Evans, and Alan~J Grodzinsky.
\newblock {Effects of short-term glucocorticoid treatment on changes in
  cartilage matrix degradation and chondrocyte gene expression induced by
  mechanical injury and inflammatory cytokines}.
\newblock {\em Arthritis research and therapy}, 13(5):R142, January 2011.

\bibitem{Natoli2009c}
R~M Natoli and K~A Athanasiou.
\newblock {Traumatic loading of articular cartilage : Mechanical and biological
  responses and post ­ injury treatment}.
\newblock {\em Biorheology}, 46(6):451--85, 2009.

\bibitem{Shampine2001}
L~F Shampine and S~Thompson.
\newblock {Solving DDEs in MATLAB}.
\newblock {\em Applied Numerical Mathematics}, 37:441--458, 2001.

\bibitem{Soetaert2012}
Karline Soetaert, Thomas Petzholdt, and R.~Woodrow Setzer.
\newblock {Package ‘deSolve’}.
\newblock Technical report, 2012.

\bibitem{Terada2011}
Chuji Terada, Aki Yoshida, Yoshihisa Nasu, Shuji Mori, Yasuko Tomono, Masato
  Tanaka, Hideo~K. Takahashi, Masahiro Nishibori, Toshifumi Ozaki, and
  Keiichiro Nishida.
\newblock {Gene expression and localization of high-mobility group box
  chromosomal protein-1 (HMGB-1) in human osteoarthritic cartilage}.
\newblock {\em Acta medica Okayama}, 65(6):369--77, December 2011.

\bibitem{Varshavsky1997}
Alexander Varshavsky.
\newblock {The N-end rule pathway of protein degradation}.
\newblock {\em Genes to cells}, 2(1):13--28, January 1997.

\bibitem{Wedlock1996}
D~Neil Wedlock, Frank~E Aldwell, and Bryce~M Buddle.
\newblock {Molecular cloning and characterization of tumor necrosis factor
  alpha (TNF-alpha) from the Australian common brushtail possum, Trichosurus
  vulpecula}.
\newblock {\em Immunology and cell biology}, 74(2):151--8, April 1996.

\bibitem{Wilson2005a}
W~Wilson, C~C van Donkelaar, R~van Rietbergen, and R~Huiskes.
\newblock {The role of computational models in the search for the mechanical
  behavior and damage mechanisms of articular cartilage}.
\newblock {\em Medical Engineering and Physics}, 27(10):810--826, 2005.

\bibitem{Zhou2004}
Shengda Zhou, Zhanfeng Cui, and Jill P~G Urban.
\newblock {Factors influencing the oxygen concentration gradient from the
  synovial surface of articular cartilage to the cartilage-bone interface: a
  modeling study}.
\newblock {\em Arthritis and Rheumatism}, 50(12):3915--24, December 2004.

\end{thebibliography}

%------FIGURES AND TABLES-----------------------------------------------------------------------------

\begin{table}\label{table:parameters}
\centering
    \begin{tabular}{|c|c|c|c|}        
         \hline
           Parameter &  Value & Units & Reason\\
         \hline
           $D_R$ & 0.1 & $\frac{\text{cm}^2}{\text{day}}$ & Determined from \cite{Leddy2004}\\
         \hline
           $D_M$ & 0.05 & $\frac{\text{cm}^2}{\text{day}}$ & Determined from \cite{Leddy2004}\\
         \hline
           $D_P$ & 0.005 & $\frac{\text{cm}^2}{\text{day}}$ & Determined from \cite{Leddy2004}\\
         \hline
           $D_F$ & 0.05 & $\frac{\text{cm}^2}{\text{day}}$ & Determined from \cite{Leddy2004}\\
         \hline
           $\delta_R$ & 60 & $\frac{1}{\text{day}}$ & Approximated\\
         \hline
           $\delta_M$ & 0.5545 & $\frac{1}{\text{day}}$ & Determined from \cite{Ito2008}\\
         \hline 
           $\delta_F$ & 0.1664 & $\frac{1}{\text{day}}$ & Determined from \cite{Wedlock1996}\\
         \hline  
           $\delta_P$ & 3.326 & $\frac{1}{\text{day}}$ & Taken from \cite{Eckardt1989}\\
         \hline     
           $\delta_U$ & 0.0193 & $\frac{1}{\text{day}}$ & Determined from \cite{Lu2011}\\
         \hline   
           $\sigma_R$ & 0.0024 & $\frac{\text{nanomolar}\cdot \text{cm}^2}{\text{day}\cdot \text{cells}}$ & Determined from \cite{Zhou2004}\\
         \hline  
           $\sigma_M$ & $5.17 \cdot 10^{-7}$ & $\frac{\text{nanomolar}\cdot \text{cm}^2}{\text{day}\cdot \text{cells}}$ & Determined from \cite{Terada2011}\\
         \hline  
           $\sigma_F$ & $2.35 \cdot 10^{-7}$ & $\frac{\text{nanomolar}\cdot \text{cm}^2}{\text{day}\cdot \text{cells}}$ & Determined from \cite{Terada2011}\\
         \hline 
           $\sigma_P$ & $4.2 \cdot 10^{-5}$ or 0.0033 & $\frac{\text{nanomolar}\cdot \text{cm}^2}{\text{day}\cdot \text{cells}}$ & Determined from \cite{Brines2008}\\
         \hline  
           $\sigma_U$ & 0.0154 & $\frac{\text{nanomolar}\cdot \text{cm}^2}{\text{day}\cdot \text{mg}}$ & Determined from \cite{Lu2011} \\
         \hline           
           $\Lambda$ & 0.5 & nanomolar & Approximated\\
         \hline                                                                         
           $\lambda_R$ & 10 & nanomolar  & Approximated\\
         \hline                                                                                  
           $\lambda_M$ & 0.5 & nanomolar  & Approximated\\
         \hline                                                                                           
           $\lambda_F$ & 0.5 & nanomolar  & Approximated\\
         \hline                                                                                           
           $\lambda_P$ & 0.5 & nanomolar & Approximated \\
         \hline                                                                                           
           $\lambda_U$ & 1 & mg/$\text{cm}^2$ & Approximated\\
         \hline                                                                                           
           $K_U$ & 0.0545 &  proportion &  Experimental setting \cite{Brouillette2013}\\
         \hline                                                                                                    
           $\alpha$ & 1 & $\frac{1}{\text{day}}$ & Approximated\\
         \hline   
           $\beta_1$ & 10 & $\frac{1}{\text{day}}$ & Approximated\\
         \hline 
           $\beta_2$ & 5 & $\frac{1}{\text{day}}$ & Approximated\\
         \hline       
           $\gamma$ & 1 & $\frac{1}{\text{day}}$ & Approximated\\
         \hline     
           $\nu$ & 0.05 & $\frac{1}{\text{day}}$ & Approximated\\
         \hline   
           $p_0$ & 1 & $\frac{1}{\text{day}}$ & Experimental setting \cite{Brouillette2013}\\
         \hline  
           $\mu_{S_A}$ & 0.1 & $\frac{1}{\text{day}}$ & Approximated\\
         \hline 
           $\mu_{D_N}$ & 0.05 & $\frac{1}{\text{day}}$ & Approximated\\ 
         \hline  
           $P_c$ & 1 & nanomolar &Taken from \cite{Brines2008}\\
         \hline  
           $\tau_1$ & 0.5 & day &Taken from \cite{Brines2008}\\
         \hline
           $\tau_2$ & 1 & day &Taken from \cite{Brines2008}\\
         \hline                                                                               
    \end{tabular}
\caption{Parameter Values}
\end{table}

\begin{table}
\centering
    \begin{tabular}{|c|c|c|c|c|c|c|}        
         \hline
           variable &  Strain=30\%  & Strain=40\%  & Strain=60\%   & Strain=80\% \\
         \hline
           C  min&  1.374 $\times 10^4$  & 2.58$\times 10^3$  & 71.02  & 0 \\
         \hline 
           $S_T$ max   & 7.254$\times 10^4$  &  8.627$\times 10^4$ & 8.909 $\times 10^4$ &  9.07 $\times 10^4$\\   
         \hline    
           $S_A$ max   & 2.79$\times 10^3$  & 5.524 $\times 10^3$  & 1.105 $\times 10^4$ & 1.533 $\times 10^4$\\  
         \hline          
           $D_A$ max   & 22   &  64  &  180  &  302\\   
         \hline  
           $D_N$ max   & 1.213 $\times 10^4$   & 2.888 $\times 10^4$  & 6.35 $\times 10^4$ & 8.616 $\times 10^4$\\   
         \hline  
           $U$ min  &29.933 & 29.894  & 29.862  & 29.824 \\ 
         \hline  
           $F$ max  & 0.0257 &  0.0332 &   0.0411 &  0.0476\\  
         \hline 
           $M$ max  &  0.0253  &  0.0428 &  0.0663 & 0.0744\\  
         \hline 
           $P$ max &  0.3234  &   0.3159 &  0.3038 & 0.3006\\ 
         \hline   
           $R$ max  &  2.7323  & 3.3031  &  3.479 & 3.493\\  
         \hline                                                                               
    \end{tabular}
\caption{Table of variable ranges under low EPO production at t=10 days}
\end{table}

\begin{table}
\centering
    \begin{tabular}{|c|c|c|c|c|c|c|}        
         \hline
           variable  & Strain=30\%   & Strain=40\%  & Strain=60\%   & Strain=80\% \\
         \hline
           C  min &8.029 $\times 10^4$  & 5.74$\times 10^4$  & 1.01 $\times 10^4$  & 0 \\
         \hline 
           $S_T$ max  & 4.778$\times 10^3$  &  7.527$\times 10^3$ & 2.78 $\times 10^4$ &  9.022 $\times 10^4$\\   
         \hline    
           $S_A$ max  & 21.29  & 38.28  & 118 & 355\\  
         \hline          
           $D_A$ max  & 0.099   &  0.39  &  1.648  &  7\\   
         \hline  
           $D_N$ max  & 1.213 $\times 10^4$   & 2.887 $\times 10^4$  & 6.35 $\times 10^4$ & 8.616 $\times 10^4$\\  
         \hline  
           $U$ min  &29.999 & 29.9988  & 29.9987  & 29.98 \\ 
         \hline  
           $F$ max  & 0.0027 &  0.0037 &   0.005 &  0.0073\\  
         \hline 
           $M$ max  &  0.0082  &  0.0168 &  0.0349 & 0.0411\\  
         \hline 
           $P$ max  &  7.919  &   9.2917 &  13.718 & 21.28\\ 
         \hline   
           $R$ max  &  0.189  & 0.2967  &  0.5139 & 0.699\\  
         \hline                                                                               
    \end{tabular}
\caption{Table of variable ranges under high EPO production at t=10 days}
\end{table}

\begin{table}\label{table:infinityLowEPO}
\centering
\vspace{2cm}
    \begin{tabular}{|c|c|c|c|c|}        
         \hline
           variable  &  Strain=30\% & Strain=40\% & Strain=60\% & Strain=80\% \\
           \hline
           C& 0.0242 & 0.0314 & 0.0244 & 0.0220 \\
           \hline
           $S_T$ & 0.0230 & 0.0225 & 0.0219 & 0.0212 \\
           \hline 
           $S_A$ & 0.0383 & 0.0352 & 0.0304 & 0.0272 \\
           \hline
           $D_A$ & 0.0525 & 0.0460 & 0.0371 & 0.0318 \\
           \hline
           $D_N$ & 0.0000 & 0.0000 & 0.0000 & 0.0000 \\
           \hline
           ECM & 0.0000& 0.0000 & 0.0000 & 0.0000 \\
           \hline
           TNF-$\alpha$ & 0.0224 & 0.0204 & 0.0204 & 0.0191 \\
           \hline
           EPO & 0.0240  & 0.0244 & 0.0319 & 0.0323 \\
           \hline
           DAMPs & 0.0219 & 0.0204 & 0.0184 & 0.0171 \\
           \hline
           ROS & 0.0230 & 0.0226 & 0.0838 & 0.1085\\
           \hline                                                                               
    \end{tabular}
\caption{$\infty$-norm relative errors under low EPO production}
\end{table}

\begin{table}\label{table:2LowEPO}
\centering
\vspace{2cm}
    \begin{tabular}{|c|c|c|c|c|}        
         \hline
           variable  &  Strain=30\% & Strain=40\% & Strain=60\% & Strain=80\% \\
           \hline
           C& 0.0102 & 0.0132 & 0.0122 & 0.0110 \\
           \hline
           $S_A$ & 0.0140 & 0.0131 & 0.01237 & 0.0118 \\
           \hline 
           $S_T$ & 0.0206 & 0.0186 & 0.0153 & 0.0134 \\
           \hline
           $D_A$ & 0.0261 & 0.0230 & 0.0180 & 0.0151 \\
           \hline
           $D_N$ & 0.0000 & 0.0000 & 0.0000 & 0.0000 \\
           \hline
           ECM & 0.0000& 0.0000 & 0.0000 & 0.0000 \\
           \hline
           TNF-$\alpha$ & 0.0148 & 0.0122 & 0.0204 & 0.0114 \\
           \hline
           EPO & 0.0134  & 0.0144 & 0.0170 & 0.0175 \\
           \hline
           DAMPs & 0.0152 & 0.0137 & 0.0118 & 0.0107 \\
           \hline
           ROS & 0.0140 & 0.0133 & 0.0166 & 0.0235\\
           \hline                                                      
    \end{tabular}
\caption{2-norm relative errors under low EPO production}
\end{table}

\pagebreak

\begin{table}\label{table:infinityHighEPO}
\centering
\vspace{2cm}
    \begin{tabular}{|c|c|c|c|c|}        
         \hline
           variable  &  Strain=30\% & Strain=40\% & Strain=60\% & Strain=80\% \\
           \hline
           C& 0.0006 & 0.0013 & 0.0045 & 0.0141 \\
           \hline
           $S_T$ & 0.0253 & 0.0305 & 0.0219 & 0.0420 \\
           \hline 
           $S_A$ & 0.0375 & 0.0362 & 0.0304 & 0.0416 \\
           \hline
           $D_A$ & 0.0430 & 0.0588 & 0.0371 & 0.0559 \\
           \hline
           $D_N$ & 0.0000 & 0.0000 & 0.0000 & 0.0000 \\
           \hline
           ECM & 0.0000& 0.0000 & 0.0000 & 0.0000 \\
           \hline
           TNF-$\alpha$ & 0.0213 & 0.0204 & 0.0203 & 0.0185 \\
           \hline
           EPO & 0.0249  & 0.0246 & 0.0306 & 0.0372 \\
           \hline
           DAMPs & 0.0188 & 0.0212 & 0.0220 & 0.0192 \\
           \hline
           ROS & 0.0253 & 0.0249 & 0.0743 & 0.1144\\
           \hline                                                                               
    \end{tabular}
\caption{$\infty$-norm relative errors under high EPO production}
\end{table}

\begin{table}\label{table:2HighEPO}
\centering
\vspace{2cm}
    \begin{tabular}{|c|c|c|c|c|}        
         \hline
           variable  &  Strain=30\% & Strain=40\% & Strain=60\% & Strain=80\% \\
           \hline
           C& 0.0002 & 0.0003 & 0.0008 & 0.0015 \\
           \hline
           $S_A$ & 0.0138 & 0.0134 & 0.0127 & 0.0116 \\
           \hline 
           $S_T$ & 0.0164 & 0.0158 & 0.0143 & 0.0119 \\
           \hline
           $D_A$ & 0.0191 & 0.0192 & 0.0200 & 0.0167 \\
           \hline
           $D_N$ & 0.0000 & 0.0000 & 0.0000 & 0.0000 \\
           \hline
           ECM & 0.0000& 0.0000 & 0.0000 & 0.0000 \\
           \hline
           TNF-$\alpha$ & 0.0128 & 0.0115 & 0.0204 & 0.0104 \\
           \hline
           EPO & 0.0130  & 0.0126 & 0.0135 & 0.0158 \\
           \hline
           DAMPs & 0.0124 & 0.0120 & 0.0115 & 0.0105 \\
           \hline
           ROS & 0.0135 & 0.0132 & 0.0148 & 0.0247\\
           \hline                                                                               
    \end{tabular}
\caption{2-norm relative errors under high EPO production}
\end{table}

\end{document}